\documentclass{ws-mpla}
\begin{document}
\markboth{Michael Grady}{TOWARD A PROOF OF LONG RANGE ORDER IN 4-D SU(N) LATTICE GAUGE THEORY}
\title{TOWARD A PROOF OF LONG RANGE ORDER IN 4-D SU(N) LATTICE GAUGE THEORY}
\author{Michael Grady\\
Department of Physics\\ State University of New York at Fredonia\\
Fredonia NY 14063 USA\\ph:(716)673-4624, fax:(716)673-3347, email: grady@fredonia.edu}
\maketitle
\begin{abstract}
An extended version of 4-d SU(2) lattice gauge theory is considered in which different inverse coupling
parameters are used, $\beta _H=4/g_{H}^2$ for plaquettes which are purely spacelike, and $\beta _V$ 
for those which
involve the Euclidean timelike direction.  It is shown that when $\beta _H = \infty$ the
partition function becomes, in the Coulomb Gauge, exactly that of a set of 
non-interacting 3-d O(4) classical Heisenberg models.  Long range order at low temperatures
(weak coupling) has been rigorously proven for this model.  It is shown that the correlation function demonstrating
spontaneous
magnetization in the ferromagnetic phase is a continuous function of $g_H$ at $g_H =0$ and therefore
that the spontaneously
broken phase enters the ($\beta _H$, $\beta _V$) phase plane (no step discontinuity at the edge).  Once the 
phase transition line has entered,
it can only exit at another identified edge, which
requires the SU(2) gauge theory within also to have a phase transition at finite $\beta$.  
A phase exhibiting spontaneous breaking 
of the remnant symmetry
left after Coulomb gauge fixing, the relevant symmetry here, is non-confining.  
Easy extension to the SU(N) case implies that the continuum limit of zero-temperature 4-d SU(N) lattice 
gauge theories is not confining,
in other words gluons by themselves do not produce a confinement.
\end{abstract}

\ccode{PACS Nos.: 11.15.Ha, 11.30.Qc, 64.60.De, 5.50.+q}

\section{Introduction}
For many years it has been believed that SU(N) lattice gauge theories exist in a single phase.  Since
confinement is a property of all compact lattice gauge theories at strong coupling, this assumption has the 
consequence that these theories also have a confining force (linear potential) in the weak coupling continuum limit.
This fits rather nicely with phenomenological evidence that the strong interactions confine quarks.  The numerical
evidence for lack of a phase transition in SU(2) comes mostly from the apparently 
smooth behavior of the 
specific heat as a function of the coupling parameter, $\beta$. However, if the critical exponent $\nu$ is 
large enough to produce a large negative specific heat exponent $\alpha$ ($\alpha = 2-d\nu $), then the first
infinite  singularity 
may be in a high derivative, and not easily visible numerically.  This is, for instance, often the case for 
3-d spin glasses which are close to their lower critical dimension.  The argument {\em in favor} of the existence 
a phase transition rests
on the similarity of lattice gauge theories to ferromagnetic spin models, all of which in three or more dimensions, at 
least for short range interactions, have two phases, with the weak coupling
phase exhibiting spontaneous symmetry breaking and long range order.  Abelian and non-abelian spin models
differ in detail, but {\em all} have phase transitions.  A major difference between gauge models and spin models are
that the former have local symmetries and the latter global.  This difference can sometimes be erased through
gauge fixing, however.  For instance the 2-d lattice gauge theory is exactly equal to a set of non-interacting
1-d classical spin chains in the axial gauge, neglecting possible boundary effects.  Setting all of the links in
one direction equal to unity using the gauge freedom causes the 4-link plaquette action to collapse to a two-link
dot product between the remaining gauge links which can be reinterpreted as spins.\cite{kogut} Another exact mapping
is between the 3-d Ising gauge theory in axial gauge with the 3-d Ising model.\cite{wegner} For 4-d theories, the Coulomb
gauge appears to be the most ``spin-like."  Here the gauge freedom is used to maximize the traces of all of the
links lying along three of the four lattice directions.  In
this gauge the fourth direction lying links become observables and act much like spins, in that they transform 
under a remnant 3-d
{\em global} symmetry on each spacelike hyperlayer that remains after Coulomb gauge fixing. This symmetry can break
spontaneously if the average of fourth-direction links over a hyperlayer acquires a magnetization.  It has been shown
that if this symmetry does break spontaneously, then the corresponding ferromagnetic phase is non-confining, so
the average of these 4-th direction links serves as a local order 
parameter for confinement.\cite{zw,gs,mp}
According to the standard hypothesis, the theory remains in the paramagnetic phase for all couplings in the
non-abelian case, and undergoes a magnetic phase transition only in the abelian case, such as U(1).  However,
below it will be shown that this inconsistent with the known long-range order for
classical 3-d Heisenberg models, to which the 4-d lattice gauge theories are closely related and actually connected in a larger
coupling space.

Numerical evidence for a ferromagnetic phase transition based on the behavior of the 
Coulomb gauge spin-like order parameter described
above has been given for SU(2) with 
the Wilson action. Finite size scaling shows an infinite-lattice
transition around $\beta = 3.2$ with a correlation length critical exponent $\nu = 1.7\pm 0.2$.\cite{cgm}  
Other simulations\cite{cgm} which 
supplement the Wilson action with an infinite chemical
potential for gauge-invariant SO(3)-Z2 monopoles\cite{so3-z2}, together 
with a positive plaquette constraint, appear to remain in the 
ferromagnetic phase for all couplings.  This would seem to indicate
that the normal confinement seen with the Wilson action is
due to strong-coupling lattice artifacts, not unlike the U(1) case.  In the following, the analytic case for
the existence of this phase transition is explored.  An analytic proof for the opposite hypothesis, namely
the lack of a phase transition
with confinement persisting to the continuum limit, was presented by Tomboulis some time ago\cite{tomboulis} 
and updated more recently.\cite{tomboulisnew}  This
proof is fairly complex and some possible flaws have been noted.\cite{seiler} When an hypothesis is difficult to prove, it is 
sometimes worthwhile to attempt to prove the opposite, which is the approach taken below.  

\section{Extended Coupling Plane}
The 4-d SU(2) theory can be extended into a larger coupling space by allowing the coupling for purely
spacelike (horizontal) plaquettes to differ from that of plaquettes which include the 
Euclidean timelike direction (vertical).  The action is 
\begin{eqnarray}
S & = & \sum _{\vec{n}} \left(\beta _H \sum _{i=1}^{2} \sum _{j=i+1}^{3} (1-\frac{1}{2} \mathrm{tr}( U_{\vec{n},i}
U_{\vec{n}+\hat{\imath},j}
U_{\vec{n}+\hat{\jmath},i}^{\dagger}U_{\vec{n},j}^{\dagger}))\right. \\
& & +\mbox{}\beta _V  \left. \sum _{i=1}^{3}  
(1-\frac{1}{2} \mathrm{tr}( U_{\vec{n},i}U_{\vec{n}+\hat{\imath},4}
U_{\vec{n}+\hat{4},i}^{\dagger}U_{\vec{n},4}^{\dagger}))\right) \nonumber 
\end{eqnarray}
Where the $U_{\vec{n},j}$ are SU(2) valued gauge links based at lattice site $\vec{n}$ in direction $j$. 
The normal Wilson action is simply
the $\beta _V = \beta _H$ case, however because the renormalization group gives a relation between $\beta$ and the
physical lattice spacing through the running coupling, the $\beta _V \ne \beta _H$ theory can also be considered
an SU(2) LGT with unequal lattice spacings in the spacelike and Euclidean timelike directions. So at least in the 
phase connected to the continuum limit (neighborhood 
of ($\beta _ H$, $\beta _ V$) $\rightarrow$ ($\infty$,$\infty$)) the entire phase plane excluding the boundaries 
can be considered an SU(2) LGT.

Consider now the $\beta _H = \infty$ theory.  This can be seen to be equivalent to a set of non-interacting 3-d
O(4) classical Heisenberg models as follows.  At $\beta _H = \infty$ all of the horizontal plaquettes will be 
forced to their largest possible value of unity.
One can then find a gauge in which all horizontal links are unity also, as follows.  Set a maximal tree of links
to unity on each spacelike hypersurface. This is a partial axial gauge in which the final trunk of the tree
along the Euclidean timelike direction is not completed. For instance for an $L^4$ lattice, on a given spacelike 
hyperlayer all 3-direction links can be set to unity except when $z=0$.  On the $z=0$ plane all 2-direction links are
set to unity except for when $y=0$ and along the line ($y=0$, $z=0$) all 1-direction links except for when $x=0$ can be set 
to unity by gauge transformations. Looking at the $z=0$ plane, there is a set of $x$-$y$  plaquettes extending backward
from $x=0$ (through the periodic boundary condition) which have three links set to unity.  The plaquette being
unity due to $\beta _H = \infty$ ensures that the fourth link in the plaquette is also unity. Now the same is true 
for the next row of plaquettes etc., forcing all links to unity except for the last row pointing in the positive direction
from x=0.  These are now equal to the gauge invariant Polyakov loop for that direction, and all of these links
are equal.  This same procedure can be extended from the ($x=0$, $y=0$) plane along the $z$ direction to
show that the links out of the plane are also unity, except for one at the edge along each lattice direction which
are equal to their neighboring links pointing in that direction.  Finally,
a Polyakov loop symmetry transformation (also a symmetry of the action) can be employed to bring
these final set of links to unity.  In this restricted sector, the Polyakov loop symmetry is the whole SU(2) group
rather than just the center. One could also formulate the theory with open boundary conditions, in which case this
final step would be unnecessary.  Although the language of axial gauge was used above, one can see that actually
the condition for Coulomb gauge has also been met, that the sum of traces of all horizontal links is maximized.  Coulomb
gauge leaves a layered remnant symmetry, one global SU(2) per hyperlayer, unfixed.  Because Coulomb gauge
was being sought is why the axial
tree above was left uncompleted. 

Since all of the horizontal plaquettes are unity, the vertical plaquettes simplify to
\begin{equation}
\frac{1}{2}\mathrm{tr}(U_{\vec{n},4}^{\dagger}U_{\vec{n}+\hat{\jmath},4})
\end{equation}
where $j$, of course, can be 1, 2 or 3.  
Writing each $U$ as
\begin{equation}
U=s_0 + i\sum _{k=1}^{3} s_k \tau _k
\end{equation}
where the $\tau _k$ are the Pauli matrices, one can associate an O(4) unit vector 
$\vec{s}=(s_0$, $s_1$, $s_2$, $s_3)$ with the SU(2) valued link.  In terms of the $\vec{s}$ 's it is easily
verified that the vertical plaquette above is simply the nearest neighbor O(4) dot product,
\begin{equation}
\frac{1}{2}\mathrm{tr}(U_{\vec{n},4}^{\dagger}U_{\vec{n}+\hat{\jmath},4})=\vec{s}_{\vec{n}}\cdot \vec{s}_{\vec{n}+\hat{\jmath}}
\end{equation}
The action on each hyperlayer is that of the 3-d O(4) Heisenberg model with coupling parameter (inverse temperature)
$\beta _V$.  In a gauge theory there are never any direct interactions between links longitudinally, and the 
freezing of horizontal links prevents any indirect interactions among hyperlayers, so these all become independent
Heisenberg models.  Each of these has its own SU(2) global symmetry, the remnant
symmetry from the Coulomb gauge.  The reason behind this is that an SU(2) gauge transformation which is {\em global}
on the
hyperlayer transforms horizontal links in such a way as to leave the trace of each link unchanged.  Thus 
the gauge condition, which is to maximize these traces, is not disturbed.  Therefore, at the level of the partition function,
the $\beta _H = \infty$ SU(2) lattice gauge theory becomes a set of non-interacting 3-d O(4) classical 
Heisenberg Models.  Durhus and Fr\"{o}hlich\cite{df} earlier pointed out a connection between SU(2) lattice gauge theory
and the 3-d O(4) model, but did not consider the case of split vertical and horizontal couplings which allows for
an exact mapping.

Long range order (LRO) in the classical Heisenberg model has been rigorously proven.\cite{lro} This means that
a ferromagnetic phase must
exist at a finite weak coupling (large $\beta _V$).  Because this is a symmetry broken phase, it must be separated 
from the strong-coupling paramagnetic phase 
by a phase transition, which has been convincingly found by Monte Carlo 
simulation at a coupling $\beta = 0.9360(1)$.\cite{o4} 
The LRO proof does not depend on the specific symmetry group.  Indeed, the existence of phase 
transitions in all ferromagnetic
spin models is well-established and non-controversial from both analytic and numerical perspectives.
The ferromagnetic order for the Heisenberg model is also quite robust to the addition other interactions.  
For instance it is still preserved at zero
temperature even 
if up to 20\% of the interactions are switched to anti-ferromagnetic (beyond this level it enters 
a spin glass phase)\cite{spinglass}, and at higher temperatures with smaller but still substantial contaminations.  
Below, 
it will be argued that the ferromagnetic phase, and 
therefore also the phase transition, continues to exist for non-infinite $\beta _H$ as well. 

In Fig.~1 the $\beta _H$ - $\beta _V$ phase plane is shown, with the 3-d O(4) model existing on the top 
edge ($\beta _H = \infty$).  The normal 4-d SU(2) model exists on the $\beta _V =\beta _H$ line.  However, 
as stated before, the $\beta _V \ne \beta _H$ cases can 
also be considered to be SU(2) lattice gauge theories, with a different
lattice spacing in the fourth direction than the other three.  Thus the entire interior, at least
in the vicinity of the continuum limit in the 
upper right corner, is an SU(2) lattice gauge theory.  

\begin{figure}[ht]
                      \includegraphics[width=3.5in]{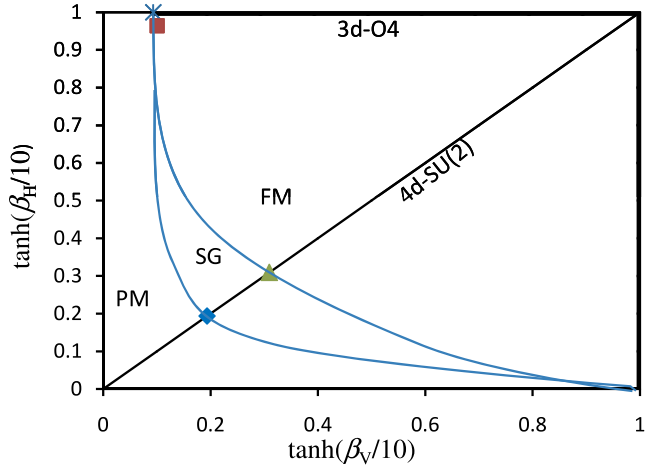}
                       \caption{Possible phase diagram for SU(2) lattice gauge theory with different horizontal and vertical couplings.
Top horizontal axis is 3-d O(4) Heisenberg model and the star shows its known ferromagnetic phase transition.\protect\cite{o4} Square
and triangle show locations of phase transitions at $\beta _H = 20$ and $\beta _ H = \beta _V$ found from Monte Carlo simulations 
using the Coulomb gauge magnetization.\protect\cite{cgm} Diamond is possible spin-glass to paramagnetic transition found previously using 
the real replica method.\protect\cite{su2sg} Bold regions of upper and right edges show portions of the border known to be ferromagnetic.
Lines drawn are hypothetical phase boundaries guided by the Monte Carlo results.\protect\label{fig1}}  
       \end{figure}
The bottom edge, with $\beta _H =0$, has only 
the vertical plaquettes left.
In an axial gauge with the fourth direction links set to unity, the system becomes a set of 1-d spin chains, which
are paramagnetic except at zero temperature ($\beta _V = \infty$) where there is a phase transition to the ferromagnetic
ground state.  The right edge of the phase diagram, $\beta _V = \infty$, is an odd phase where all of the vertical 
plaquettes are forced to unity. Staying with the axial gauge, one can see that the 3-d SU(2) theories on spacelike
hypersurfaces but different Euclidean times are {\em locked} to each other, 
in other words they can only fluctuate 
in lockstep.  This makes for a theory with a 4-d energy, but non-extensive 3-d entropy, so it is basically
stuck in the classical ground state. Both top and bottom ends of this line are ferromagnetic; there is no reason 
for the whole line not to be also.  Thus ferromagnetism exists on the upper border for 
$\beta _V > 0.936$ and along the entire right-side border.
In the following, two arguments are given which show that the ferromagnetic phase persists
as one enters the interior of the phase diagram. An important point is that the global SU(2) symmetries
of the independent Heisenberg models persist for the $\beta _H < \infty$ case in the form of the Coulomb Gauge 
remnant symmetries
which exist independently on each hyperlayer.
For a symmetry
breaking phase transition, once the phase-transition line has entered the phase plane it must continue to another edge.  This is
because the order parameter, due to the realized symmetry, is exactly zero in the 
paramagnetic region on the infinite lattice, and of course
is nonzero in the ferromagnetic phase.  An analytic function zero in a finite region  is zero everywhere, so a
line of non-analyticity must separate the two phases completely.  The only place where the phase transition line
could terminate is at the only other phase transition on the border, namely the lower right corner as shown. Because 
the remnant symmetry exists for both the $\beta _H = \infty$ and the $\beta _H$ finite cases, this situation is not
analogous to the Ising model in an external field, where the field, no matter how small, removes the transition due
to explicit symmetry breaking.

Fig.~1 also shows the ferromagnetic phase transitions found with Monte Carlo simulations using the Coulomb Gauge
magnetization as the order parameter. One of these was for $\beta _H = 20$ where the phase transition on the infinite lattice
was determined to be $\beta _{Vc} = 1.01(2)$.  This transition closely resembles that of the Heisenberg model in that it has
similar critical exponents.  Finite size scaling fits are good, which indicates lattices are large enough to
suppress non-leading  effects.  Observation of such a transition in Monte Carlo simulations
is a strong indication that the phase transition does enter the 
interior of the phase diagram. Below, it will be argued that that must be the case from an analytic perspective as well.

\section{Persistence of Phase Transition for $\beta _H < \infty$}
In order for the SU(2) lattice gauge theory in the interior of the phase diagram to avoid the phase transition
as in the conventional hypothesis, the
transition
would need to be somehow prevented from entering the interior of the phase diagram at all.  This means that the 
small terms that arise in the Hamiltonian when $\beta _H$ is backed off from $\infty$ would have to destroy the 
ferromagnetic order, no matter how small the coupling $g_H$ ($\beta _H = 4/g_{H}^{2}$). This seems odd considering
the continued existence of the symmetry and the known robustness of ferromagnetism in the Heisenberg model.  
Say we start in the deep-ferromagnetic region of the Heisenberg model. 
For the order parameter to jump from a finite value at $g_H =0$ to a value of zero for any $g_H >0$ there
would have to be a first-order phase transition at the edge of the phase diagram.  However, this edge is 
already well characterized, since it is the ferromagnetic phase of the Heisenberg model itself. Known properties
of this phase are inconsistent with it being the location of a first-order phase transition 
in the {\em same order parameter} due to the following.  A symmetry-breaking first-order phase transition 
is of the type associated with
a tricritical point, which is controlled by a sixth-order Landau effective potential as shown in Fig.~2 (solid line) at
the point of phase transition.\cite{cl} For $g_H >0$ the two side-minima would hypothetically 
lift up, leaving only the
minimum at zero order parameter.  On the phase transition point, both the phase with zero order 
parameter {\em and} the two
instances of the broken-symmetry phase exist in equilibrium.  Such phase mixing would be easily observable through
an un-sharp order parameter and a latent heat (range of internal energies). However,
such a state simply does not exist in the Heisenberg model. In the deep ferromagnetic region its Landau
effective potential is widely believed to look like the dashed line in Fig.~2 with only two minima 
which is incompatible with the tricritical behavior.  
\begin{figure}[ht]
                      \includegraphics[width=3.5in]{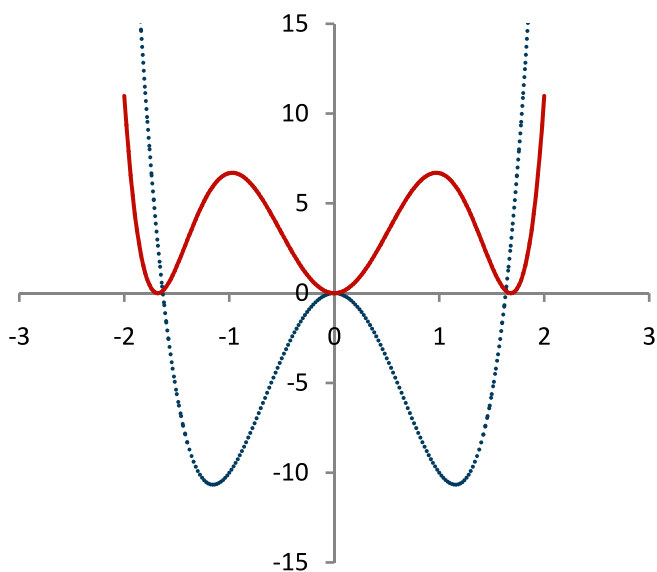}
                                  \caption{Landau effective potentials at a symmetry breaking first-order
transition (solid line) and in the deep-ferromagnetic region of a system with a 
higher-order transition (dotted line). Scales are arbitrary.}
          \label{fig2}
       \end{figure}

Therefore, the hypothesis that the ferromagnetic phase does not enter the interior of the phase diagram seems to
require behavior of the Heisenberg model inconsistent with known behavior.  If a first-order phase transition is present
it cannot be a conventional one.  However, the argument above is somewhat
heuristic and non-rigorous in that it involves  Landau effective potentials.  A more rigorous proof 
of the impossibility of a first-order phase transition as one enters the
phase plane from the deep-ferromagnetic region of the Heisenberg model can be constructed as follows.  Consider the
correlation function $C(\vec{q},\vec{r})=<\frac{1}{2}\mathrm{tr}(U_{\vec{q},4}U_{\vec{r},4}^{\dagger})>$. In the ferromagnetic
phase this shows spontaneous symmetry breaking through 
\begin{equation}
\lim _{|\vec{q}-\vec{r}|\rightarrow \infty} C(\vec{q},\vec{r})=<|\vec{m}|>^2
\end{equation}
where
\begin{equation}
\vec{m}=\frac{1}{V}\sum _{\vec{n}}\vec{s}_{\vec{n}}
\end{equation}
where the sum is over individual 3-d hypersurfaces and $V$ is the 3-volume.  The expectation value above includes an
average over different Euclidean times as well as gauge configurations.  
Although a finite lattice is used for defining quantities, the infinite lattice case is 
implicitly being considered here.

Now let us calculate the derivative of this correlation function  with respect to $g _H$ at $g _H = 0$.
Note that the calculation is performed in the Heisenberg model itself (earlier shown to be the $g _H = 0$ limit), 
but in 
order to determine what function
to calculate the expectation value of, we must consider the $g _H \ne 0$ theory.  To do this, rewrite the
horizontal $U$ links in terms of gauge fields $A$, as in the usual calculation of the continuum limit:

\begin{eqnarray}
S & = & \sum _{\vec{n}}\left( \frac{4}{g_{H}^{2}}\sum _{i,j>i}(1-\frac{1}{2}\mathrm{tr}( \exp (ig_H \vec{A}_{\vec{n},i}\cdot \vec{\tau}) 
\exp (ig_H \vec{A}_{\vec{n}+\hat{\imath},j}\cdot \vec{\tau})\right.\nonumber\\ 
& & \;\;\;\;\;\;\;\;\;\;\;\; \exp (-ig_H \vec{A}_{\vec{n}+\hat{\jmath},i}\cdot \vec{\tau}) 
\exp (-ig_H \vec{A}_{\vec{n},j}\cdot \vec{\tau}))) \nonumber\\
& & \!\!\!\! +\mbox{}\beta _V \! \! \left. \sum _{i<4}
(\! 1\! -\! \frac{1}{2}\mathrm{tr}( \exp (ig_H \vec{A}_{\vec{n},i}\cdot \! \vec{\tau}) 
U_{\vec{n}+\hat{\imath},4} \exp (-ig_H \vec{A}_{\vec{n}+\hat{4},i}\cdot \! \vec{\tau})
U_{\vec{n},4}^{\dagger}))\! \! \right)
\end{eqnarray}
where $i$,$j$ run from 1 to 3. No approximation has been made.  For finite $g _H$ this is not a practical decomposition,
since the A integrations still need to obey the Harr measure for the $U$'s, however it is useful  in the limit $g _H \rightarrow 0$. 
Taking the limit $g _H \rightarrow 0$ gives
\begin{equation}
\label{action0}
S=\sum _{\vec{n}}\left( \sum _{i,j>i}F_{ij}^2+\beta _V \sum _{i}
(1-\frac{1}{2}\mathrm{tr}(U_{\vec{n}+\hat{\imath},4} 
U_{\vec{n},4}^{\dagger}))\right)
\end{equation}
where $F_{ij}$ is the abelian field strength tensor (the non-abelian part having a factor of $g _H $).
This gives the multiple 3-d Heisenberg models as expected, but also a disconnected
set of 3-d free field theories. 
The connected part of 
\begin{equation}
\left. \frac{\partial C(\vec{q},\vec{r})}{\partial g_ H }\right| _{g _H =0}
\end{equation}
is given by 
\begin{equation}
\frac{1}{4}\beta _V \sum _{\vec{n},i<4}<\mathrm{tr}(U_{\vec{q},4}U_{\vec{r},4}^{\dagger})\mathrm{tr}((i\vec{A}_{\vec{n},i}\cdot \vec{\tau}) 
U_{\vec{n}+\hat{\imath},4}U_{\vec{n},4}^{\dagger})>.  
\end{equation}
and a similar term with the $\vec{A}$ in the other position.
This is to be computed in the $g_H = 0$ action of Eq. \ref{action0} which is even in the transformation $\vec{A}\rightarrow -\vec{A}$.
As a consequence, any expectation value containing a single factor of $\vec{A}$ is zero. Disconnected parts similarly vanish or cancel.
This almost trivial 
observation means that the first derivative of the order parameter with respect to $g_V$ at $g_V=0$ vanishes,
and therefore exists.  If the first derivative exists at a point, it follows from an elementary theorem of analysis that
the function itself is continuous at that point.\cite{rudin}  However, if the order parameter is a continuous
function of $g_V$ at $g_V =0$ then it cannot drop suddenly to zero here in a first-order phase transition, which 
would require a discontinuity.  Therefore, the ferromagnetic phase must enter the phase diagram.  In fact, since $g _H$
does not affect the phase transition at lowest order, the line would be expected to enter at a $90 ^{\circ}$ angle.

Since, as argued before, the entire interior of the phase diagram is an SU(2) LGT if unequal lattice spacings are
considered, this means that there must be a ferromagnetic phase in SU(2) as well. The SU(2) LGT has long-range order.
The order parameter is the Coulomb-gauge magnetization.  
As argued before, due to the symmetry breaking nature of the phase transition, the phase
transition line must continue to another edge, the only possibility being the lower right corner of Fig.~1.  
In doing so it clearly 
must cross the $\beta _V = \beta _H$ line, which is the ordinary SU(2) LGT with equal lattice spacings.
A phase transition in which the Coulomb Gauge remnant symmetry breaks spontaneously is known to be deconfining.\cite{zw,gs,mp}
Therefore the zero physical temperature (infinite lattice) 4-d SU(2) lattice gauge theory 
must have a deconfining phase transition, contrary to the usual assumption. To prove for SU(3) or SU(N) one only has
to replace the O(4) Heisenberg model with the SU(N)$\times$SU(N) spin model.  Since these all have ferromagnetic phases,
the argument goes through in the same way for them.

This argument has similarities to another approach which had the same conclusion.\cite{fa}  
In the fundamental-adjoint plane of 
SU(2) lattice gauge theory with couplings $\beta _F$ and $\beta _A$, there is 
a well-known first-order phase transition which starts at the 4-d Z2 LGT transition
(at $\beta _A = \infty$) and ends in the middle of the diagram at ($\beta _F$, $\beta _A$) = (1.48, 0.90).\cite{bc}  This 
has been seen as a critical point, below which an analytic path exists between the strong-coupling confining region
with the weak-coupling continuum limit.  However, that would require the first-order phase transition to be 
a non-symmetry-breaking transition.  If it were symmetry-breaking, on the other hand, then the end of the first-order line
would be a tricritical point, and the transition would have to continue as a higher-order one, bisecting the 
entire coupling plane into symmetry-broken and unbroken sectors, as above.  Interestingly, 
a critical point and a tricritical point are rather easily distinguished through the scaling behavior of the
attached first-order transition. In particular, the scaling relationship between the latent heat and the size
of the metastability region is linear in the critical case and quadratic for the tricritical case.\cite{fa}  
One simply monitors the shape of the growing hysteresis rectangle in 
the energy-coupling plane
while moving up the first-order line. 
The numerical evidence points convincingly to the tricritical case.\cite{fa} This again implies a symmetry-breaking phase transition 
separates the strong and weak-coupling regions.  One of the strengths of this energy-scaling argument is that 
it does 
not require identification of 
the symmetry or order parameter involved, and no gauge-fixing is required. 
The tricritical case simply requires {\em some} symmetry to break.
However, it seems likely it is the same symmetry breaking as studied above, something relatively easy to check.

\section{Artifact Driven Transition}
It is interesting to consider possible mechanisms which could drive the SU(2) transition.  In the U(1) case, confinement
arises from abelian monopoles, which are strong-coupling lattice artifacts.\cite{dgts}
The confined phase occurs when the 
monopoles form percolating loops.  Confinement can be prevented by suppressing monopoles with a 
chemical potential.\cite{mitrj}
Some time ago a gauge invariant monopole was introduced which could play the same role for the SU(2) 
theory.\cite{so3-z2} It carries SO(3) and Z2 degrees of freedom which in some sense cancel each other, so
was named an SO(3)-Z2 monopole.
Suppressing these monopoles, together with a positive plaquette restriction, appears to prevent the transition
to the confining phase.\cite{cgm}  The system stays in the spontaneously
magnetized phase all the way to zero coupling when this constraint is imposed.\cite{toap} Lattices to $60^4$
have been measured and the lattice spacing (determined from the running coupling) is such 
that these should definitely be in the confining region
if universality applies.  Potentials, measured to 1.5 fm, show no evidence of a linear term.
\begin{figure}[ht]
                      \includegraphics[width=3.5in]{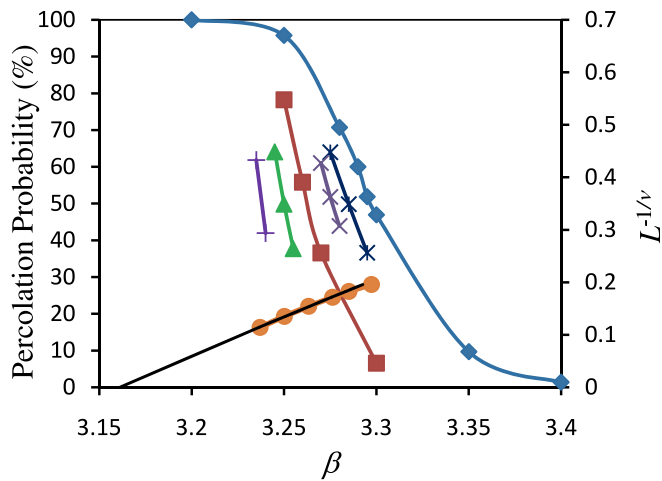}
                                  \caption{Percolation probabilities for SO(3)-Z2 monopoles as a function of $\beta$ for
various lattice sizes (diamond $16^4$, * $18^4$, $\times$ $20^4$, box $24^4$, triangle $30^4$, + $40^4$). Also shown is extrapolation
of 50\% point to the infinite lattice (uses right y-axis).}
          \label{fig3}
       \end{figure} 

It is 
interesting to monitor these monopoles
in the standard Wilson theory on periodic lattices.  They form percolating clusters in the crossover region
where confinement occurs and de-percolate at weaker couplings.  Fig.~3 shows the percentages of percolating lattices
as a function of $\beta$ for various lattice sizes.  The finite-lattice 
percolation transition can be taken to occur at the 50\% level. Also shown is the extrapolation of the percolation
transition to the infinite lattice.
The extrapolation shown uses the finite-lattice shift equation for a higher-order phase transition\cite{barber}
\begin{equation}
1/\beta _{cL}= 1/\beta _{c}+cL^{-1/\nu}
\end{equation}
using the expected value $\nu = 1.7$ from Ref.~6. Here $L$ is the linear lattice size and $\beta _{cL}$ 
is the apparent critical point on the finite lattice. A fit to the data for $c$ and $\beta _c$ gives
$\beta _c = 3.161(2)$.  If instead, $\nu$ is assumed to be unknown and determined from the fit then the best fit is 
obtained for $\nu = 1.1$ yielding $\beta _c = 3.193(1)$.  
Setting $\nu = 0.5$ gives $\beta _c = 3.226(2)$. The uncertainties in $\beta_c$
from the fixed-$\nu$ fits are much smaller than the differences in fits with different $\nu$'s, so
the majority of uncertainty in $\beta _c$ is from the extrapolation.  Allowing a broad
range for this exponent from 0.5 to 1.7 gives for the infinite lattice $\beta _c = 3.19(3)$.
It is very interesting that this is
consistent with the position of the deconfining Coulomb-gauge magnetization transition determined from  
data-collapse fits to scaling
behavior, which yielded an infinite lattice critical point of $3.18(8)$.\cite{cgm} 
It seems unlikely that these agree
by mere coincidence. The percolation study was done after the other study was completed and released.
Being that the percolation study was performed on the standard Wilson theory using conventional heat-bath and 
over-relaxation updates and with periodic
boundary conditions, the coincidence of these results for $\beta _c$ 
gives additional confidence that the open boundary-condition Coulomb-gauge methods used
in Ref.~6 are reliable.  The fact that the percolation transition moves to {\em smaller} $\beta$ as the lattice size 
is increased means it almost certainly exists on the infinite lattice.  The connection between SO(3)-Z2 monopole
percolation and confinement will be further explored in a forthcoming publication\cite{toap}.

\section {Conclusion} It has long been known that lattice gauge theories have a lot in common 
with spin models.  This becomes especially
apparent in certain fixed gauges for which the remnant symmetry can be matched onto a spin model. 
The result presented above implies spin and gauge theories
have even more in common,
namely long range ferromagnetic order.  This has important consequences for 
the non-abelian lattice gauge theories
SU(2) and SU(3) which were previously assumed to be exceptions to this behavior.  The relatively large critical
exponent, $\nu \sim 1.7$ for SU(2), explains how such a transition could have been missed, since the corresponding 
specific heat singularity is very soft.

The ordered continuum limit means that the quenched 
lattice gauge theory without fermions does not confine in the continuum limit, an idea which 
has been suggested previously.\cite{previous,ps}  
Because good phenomenological evidence exists for quark
confinement, however,
some source of confinement
must be sought, which need not necessarily be a linear potential.  
If light quarks are added to the theory, chiral symmetry is still expected to break spontaneously 
through the formation of a quark condensate.  This only requires a sufficiently strong force, 
not necessarily a confining one.
If this collective state polarizes in such a way as to repel strong color fields, the chiral condensate could form
a kind of bag surrounding mesons and baryons, contributing a confining-like term to the force over a limited range, 
though confinement
would not be absolute.\cite{csbqc,ch,gribov}  Also, if a quark were to find itself a long distance from its partner antiquark, a 
quark/antiquark current in the chiral condensate could quickly generate local partners.  This is somewhat different
from generating quark anti-quark pairs from gluon ``sparking" which originate at the same location, but that too 
can prevent isolated quarks from existing provided there is sufficient energy in the bond.  Thermodynamics of the
quark-gluon plasma is also modified in this picture, since confinement is no longer in the gluon sector.  A phase
transition or at least a rapid crossover to a chiral-symmetric phase at high physical temperatures 
will undoubtedly still exist, which, 
if chiral symmetry breaking is related
to confinement as above, will also be deconfining.

\end{document}